\documentclass[prd,a4paper,nofootinbib,showpacs
]{revtex4}
\usepackage{graphicx}
\usepackage{amsfonts}
\usepackage{color}
\def\eq#1{{eq.~(\ref{#1})}}
\def\eqs#1#2{{eqs.~(\ref{#1})--(\ref{#2})}}

\def\hbar{\hspace{0pt}\raisebox{1pt}{$-$} \hspace{-7pt} h}

\def\5{\overline 5}

\newcommand{\be}{\begin{equation}}
\newcommand{\ee}{\end{equation}}
\newcommand{\bea}{\begin{eqnarray}}
\newcommand{\eea}{\end{eqnarray}}
\newcommand{\nn}{\nonumber}

\newcommand{\ba}{\begin{eqnarray}}
\newcommand{\ea}{\end{eqnarray}}

\begin{document}
\title[]{Fermions and Goldstone bosons in an asymptotically safe  model

}

\author{F. Bazzocchi$^{\dag\ddag}$}
\author{M. Fabbrichesi$^{\ddag}$}
\author{R. Percacci$^{\dag\ddag}$}
\author{A. Tonero$^{\dag\ddag}$}
\author{L. Vecchi$^{*}$}
\affiliation{$^{\ddag}$INFN, Sezione di Trieste} 
\affiliation{$^{\dag}$SISSA,
via Bonomea 265, 34136 Trieste, Italy}
\affiliation{$^{*}$LANL, Los Alamos, NM 87545, USA}

\begin{abstract}
We consider a model in which Goldstone bosons, described by a $SU(N)$ chiral nonlinear $\sigma$ model, are coupled to an $N$-plet of colored fermions by means of a Yukawa interaction. We study the one-loop renormalization group flow and show that  
the non-Gaussian UV fixed point, which is present
in the purely bosonic model, is lost because of fermion loop effects
unless $N$ is sufficiently large. 
We then add four-fermion contact interactions to the lagrangian and show that in this case there exist several non-Gaussian fixed points. 
The strength of the contact interactions,  predicted by the requirement that the theory flows towards a fixed point in the UV, is compared to the current experimental bounds. This toy model could provide an important building block of an asymptotically safe model of the weak interactions.
\end{abstract}

\maketitle
%
\section{Motivations}

Although the  nonlinear $\sigma$ model (NL$\sigma$M)
is usually seen---because of its perturbative nonrenormalizability---as a mere low-energy effective field theory, it could also be renormalizable in a nonperturbative sense,
namely at a nontrivial (non-Gaussian) fixed point (FP) of the renormalization group (RG) flow \cite{Weinberg} (see \cite{reviews} for the application of this idea in the context of gravity); in this case, the model is said to be asymptotically safe (AS).

As appealing as this approach may seem, it has the  disadvantage that perturbation theory can only be  a rough guide, and therefore many results are necessarily only suggestive rather than conclusive.
If in spite of this serious drawback one considers the RG one-loop $\beta$-functions,
or some resummation thereof, it is easy to see that a nontrivial FP is indeed present in the NL$\sigma$M~\cite{zinn,codello}. It persists when one couples the Goldstone bosons to gauge fields \cite{fptz} and, in some cases, when one adds terms with four derivatives of the Goldstone bosons \cite{hasenfratz}. While these results do not prove the existence of the FP, they are suggestive and we use them to justify our rather phenomenological approach in which the one-loop results are assumed to hold at least qualitatively in the non-perturbative solution and the consequences of such an assumption are then worked out.

Whether the physical system described by the NL$\sigma$M is actually on a RG trajectory leading to such a FP or not depends on the initial conditions at some given energy scale
and the experimental constraints on the coefficients in the operator expansion. 
 Whereas it appears that this is not the case for the chiral NL$\sigma$M of strong interactions (for which the Goldstone bosons are the pions), this is still an open question in the case of the Goldstone bosons of the spontaneous breaking of electroweak interactions. This question will eventually be answered by the experimental data. 

In this work we address a more specific problem and reconsider the chiral NL$\sigma$M to study what happens to the FP of Ref.~\cite{codello} when the Goldstone bosons are coupled to fermions. This is an important question in view of utilizing the model in a realistic theory of the weak interactions. 

As we shall see, a one-loop computation shows that the inclusion of fermion fields drastically modifies the asymptotic properties of the NL$\sigma$M. The Yukawa coupling does not have a UV fixed point and instead grows with the energy. The Goldstone boson coupling hovers about the FP of the uncoupled model---as long as the Yukawa coupling size is small enough to be negligible---but is eventually pushed away by the increasingly large  Yukawa coupling.

We interpret this instability as a signal of new  physics associated to the standard model (SM) fermions, and in particular the top quark, and model it by means of  four-fermion contact interactions---somewhat along the lines of what is done in nuclear physics in describing the nucleon potential. The fermionic content of the model is left unchanged and identical to that of the SM. The model thus extended admits a number of non-trivial FPs and can thus be used in modeling the interaction between Goldstone bosons and fermions in a AS theory. 

We also show how current  experimental bounds on quark contact interactions already limit the choice of RG trajectories which can be used. As the bounds are improved, or the new interactions are discovered, they will provide a direct test of the model.

\section{Fermions and Goldstone bosons}

We consider a $SU(N)$--valued scalar field $U=\mathrm{exp}(i f\pi^a T_a)$
where $\pi^a$ are the Goldstone boson fields, $\mathrm{tr} T_a T_b = \delta_{ab}/2$ and $f$ is the Goldstone boson coupling, which in the SM case can be identified with $2/\upsilon$,
where $\upsilon = 246$\,GeV is the Higgs VEV.
We limit ourselves to the lowest order term in the NL$\sigma$M lagrangian, which reads
\be
\label{lag}
\mathcal{L}_\sigma=
-\frac{1}{f^2} \mathrm{Tr}
\left(U^\dagger\partial_\mu U U^\dagger\partial^\mu U\right)\, .
\ee
This model is invariant under separate $SU(N)_L$ and $SU(N)_R$ transformations
acting on $U$ by left- and right multiplication respectively.

We couple the Goldstone bosons to left- and right-handed fermions 
$\psi_L^{ia}$ and $\psi_R^{ia}$ carrying the fundamental representation of 
$SU(N)_L$ and $SU(N)_R$ respectively (corresponding to the indices $i=1,\ldots,N$), 
and also the fundamental representation of a color group $SU(N_c)$
(corresponding to the indices $a=1,\ldots,N_c$).
In the real world the latter group is gauged; here we merely retain it
as a global symmetry to count fermionic states. 
We couple the fermions in a chiral invariant way to the $U$ field by  adding to the NL$\sigma$M lagrangian the fermion kinetic  and the Yukawa terms:
\be
\label{lag2}
\mathcal{L}_{\psi^2}  = \bar{\psi}_Li\gamma^\mu \partial_\mu\psi_L+ 
\bar{\psi}_Ri\gamma^\mu \partial_\mu\psi_R 
 - \frac{2h}{f} \big( \bar \psi_L^{ia} U^{ij} \psi_R^{ja} + \mathrm{h.c.}\big) \, ,
\ee
where we have explicitly written out the group indices in the interaction.

In a more realistic model, the Yukawa coupling should distinguish among the fermion components and there should be more than one family,  as it is the case in the SM, but this is not essential for the present discussion. 
Indeed, we find that the only phenomenologically relevant fermionic
contribution comes from the top quark, so $h$ in~(\ref{lag2}) will be viewed as the top Yukawa coupling.
We also neglect the strong, weak and electromagnetic gauge couplings (after having checked that they do not alter our conclusions) and for this reason the derivatives in \eqs{lag}{lag2} are not covariant. 

\begin{figure}[h!]
\begin{center}
\includegraphics[width=4in]{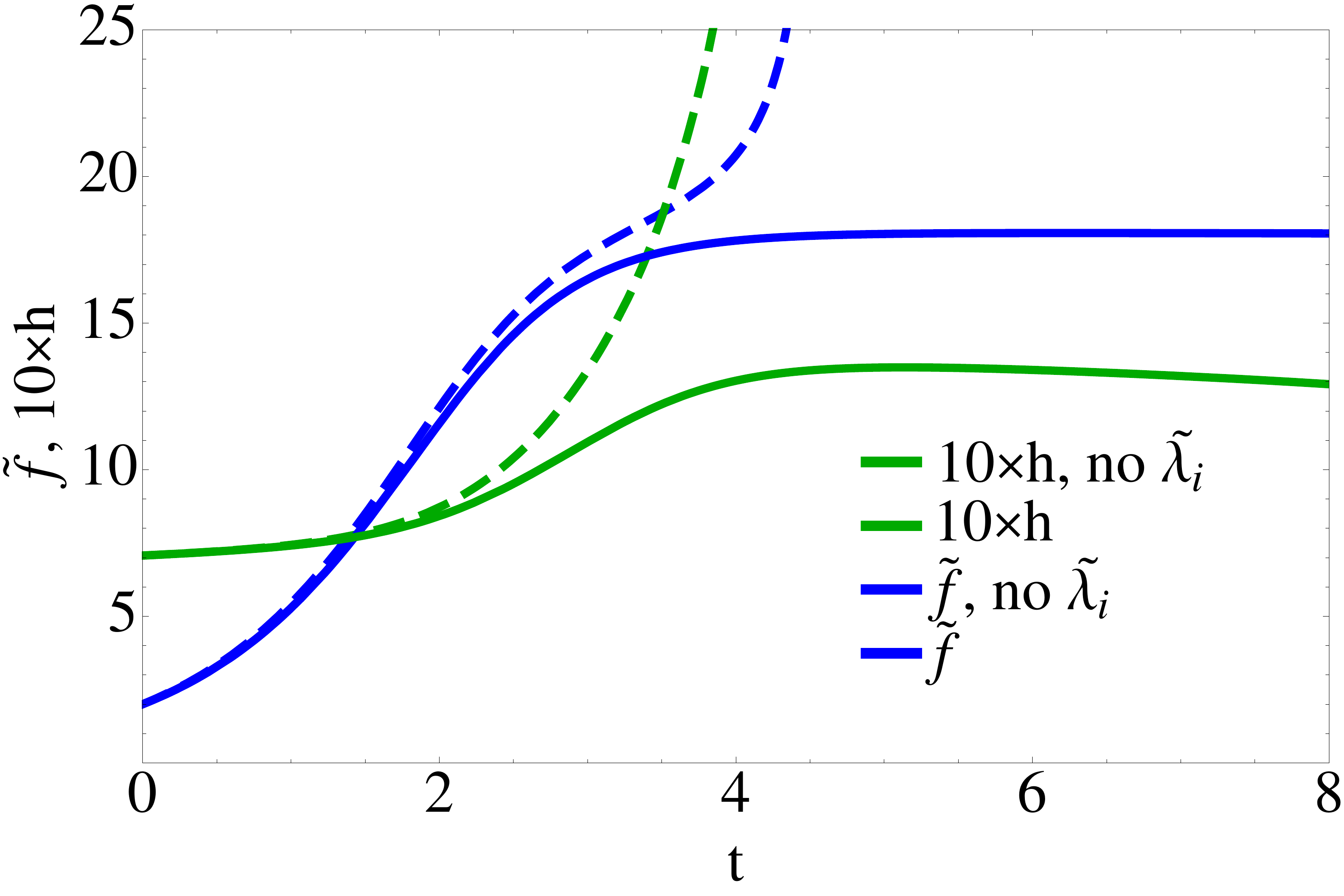} 
\caption{\small Running of $\tilde f$ and $h$ (rescaled by a factor of 10 to fit the figure) for $N=2$, $N_c=3$. 
The AS behaviour of $\tilde f$ (represented in the figure by the blue dashed  line) is destabilized around $t=4.5$ (that is, around 22 TeV) due to the increasingly large Yukawa coupling contribution (green dashed line) which at about the same value becomes strongly coupled, that is larger than $2\pi$.  The asymptotically safe behavior (continuous lines) is recovered after the introduction of the four-fermion contact interactions, as discussed in section III. \label{fermions}}
\end{center}
\end{figure}

Using a sharp cutoff regularization, we obtain the following one-loop RG equations for 
$\tilde f=k f$ and $h$:
\bea
\label{beta1}
\frac{d\tilde f}{dt}&=&
\tilde f
-\frac{N}{64\pi^2}\tilde f^3
+\frac{N_c}{4\pi^2}h^2\tilde f\ ,
\\
\label{beta1bis}
\frac{d h}{d t} & = & \frac{1}{16\pi^2} \left(4N_c- 2 \frac{N^2 -1}{N} \right)h^3 
+\frac{1}{64\pi^2}\frac{N^2 -2}{N}h\tilde f^2
\eea
The variable $t$ is defined as $\log k/k_0$ where $k$ is the renormalization group
energy scale and $k_0$ is a conventional reference energy, which we take to be $k_0=\upsilon$.

The system of equations~(\ref{beta1})-(\ref{beta1bis}) admits a number of possible UV fixed points. The Gaussian FP is the solution $\tilde f=0$, $h=0$. 
 This FP has one irrelevant (IR attractive) direction $\tilde f$ and one marginal direction $h$. Any RG trajectory starting from it in the UV would be trivial at any scale. We therefore reject this choice because it is uninteresting. We emphasize that conventional chiral perturbation theory is the study of small deformations of the FP $\tilde f=0$, interpreted as an IR FP.

There is also a nontrivial FP at $h_*=0$, $\tilde f_*=8\pi/\sqrt{N}$
for which $\tilde f$ is a relevant (UV attractive) direction and $h$ is marginally irrelevant.
Again, interpreting such a FP as an UV attractive FP would imply the triviality of the Yukawa coupling at all scales. We thus see that this FP is the same as in the purely bosonic theory \cite{codello}.

 A physically viable UV limit for~(\ref{beta1})-(\ref{beta1bis}) requires
the existence of aFP with both $h$ and $\tilde f$ being nonvanishing.
If $h$ is treated as a $t$-independent constant, the $\beta$-function for $\tilde f$
has a zero at $\tilde f_*=4\sqrt{(4\pi^2+N_c h^2)/N}$, which is a deformation of the one 
appearing in the pure bosonic model.
The existence of a nontrivial FP for the coupled system thus hinges on the existence of a nontrivial zero in the $\beta$-function of $h$.
This requires that the first term in the right hand side of \eq{beta1bis} be negative,
which is true for $N>2N_c$.

Unfortunately this condition is not satisfied
for the phenomenologically most important case $N=2$, $N_c=3$.
This is illustrated by the dashed curves in Fig.~\ref{fermions}, for the initial condition $\tilde f_0=2$ and $h_0=0.7$ at $t=0$ (thus mimicking the top-quark value).
The first term on the right hand side of \eq{beta1} is initially dominant, leading to linear
growth of $\tilde f$. The second term then grows in absolute value
and at some point nearly balances the first one, leading to an approximate
FP behavior in some range of energies. 
Eventually $h$, whose $\beta$-function is everywhere positive, becomes
large and the third term dominates leading to a Landau pole and the loss of AS.
The scale at which destabilization occurs  is very sensitive to the initial conditions and for the Yukawa couplings corresponding to light fermions
no destabilization takes place up to very large energies. 

We conclude that our model is not AS in the case $N=2$, $N_c=3$, in the one loop approximation.
For it to be AS, either the one loop approximation must break down,
or else new physical effects must enter in the fermion sector at some  energy scale.

It is interesting to compare this behavior to similar models.

If the color symmetry was gauged there would be an additional contribution
to \eq{beta1bis}:
\be
-\frac{3}{16\pi^2}\frac{N_c^2-1}{N_c} g_s^2 h  \label{A}
\ee
The system~(\ref{beta1})-(\ref{beta1bis}) would then resemble the SM where the second term in \eq{beta1bis} is abstent. This model is found to predict a perturbative Yukawa coupling up to very high scales, in agreement with~\cite{ramond}. On the other hand, in the case of the NL$\sigma$M the second term in \eq{beta1bis} is present and  the qualitative behavior shown in Fig.~\ref{fermions} would not be changed by the term~(\ref{A}).

A strictly related model is the linear sigma model coupled to one right-handed and $N_L$ left-handed fermions, studied in \cite{Gies}.
Our Goldstone modes are contained in their scalar sector, with the
VEV $\upsilon=2/f$ corresponding to the minimum of the scalar potential.
There it was found that the the scalar potential and the Yukawa coupling
admit a FP for $1\leq N_L\leq 57$.
Our results are quite different due both to the different fermion content and to the non-linear boson-fermion coupling.

\section{Four-fermion interactions}

New physics associated to the SM fermions might restore asymptotic safety. In this section we will discuss a class of nontrivial UV FPs with asymptotically free Yukawa couplings that emerge once short-range interactions among the fermions are included.

We now restrict ourselves to the case $N=2$ and add to the lagrangian  a complete set of $SU(2)_L\times SU(2)_R$ invariant four fermion operators.
Requiring $P$ invariance, all possible chiral invariant operators,
up to Fierz reorderings are given by the following lagrangian:
\bea\label{4f}
{\cal L}_{\psi^4} &=& \lambda_{1} \, \left( \bar \psi_L^{ia} \psi_R^{ja} \bar \psi_R^{jb} \psi_L^{ib}\right)+ \lambda_{2} \, \left(\bar \psi_L^{ia} \psi_R^{jb}  \bar \psi_R^{jb} \psi_L^{ia} \right)\nn\\
&+& \lambda_{3} \, \left( \bar \psi_L^{ia}\gamma_\mu \psi_L^{ia} \bar \psi_L^{jb}\gamma^\mu \psi_L^{jb} +\bar \psi_R^{ia}\gamma_\mu \psi_R^{ia} \bar \psi_R^{jb}\gamma^\mu \psi_R^{jb}\right)\nn\\
&+& \lambda_{4} \, \left( \bar \psi_L^{ia}\gamma_\mu \psi_L^{ib} \bar \psi_L^{jb}\gamma^\mu \psi_L^{ja} +\bar \psi_R^{ia}\gamma_\mu \psi_R^{ib} \bar \psi_R^{jb}\gamma^\mu \psi_R^{ja}\right) \, .
\eea
 The coefficients $\lambda_i$  have inverse square mass dimension.
We do not include in the lagrangian in \eq{4f} operators defined by taking the square of the Yukawa term in \eq{lag2} because they are higher order from the point of view of chiral perturbation theory.

Strictly speaking, only the third SM fermion generation requires new physics to emerge at relatively low scales, since four-fermion operators involving the first two generations can be suppressed by much larger scales without spoiling our AS scenario. Yet, in discussing the experimental bounds on our model, see Section IV, we will be as conservative as possible and consider the rather pessimistic scenario in which~(\ref{4f}) also involve the first generation. We thus tacitly assume that the operators in~(\ref{4f}) are consistent with FCNC bounds.

The symmetries imposed on \eq{4f} make this lagrangian the minimal choice, and the one we will study here. More general sets of operators may well be relevant depending on the symmetries of this new sector;  the SM group $SU(2)\times U(1)$ being the first instance coming to mind. The analysis is in this case more cumbersome because more operators must be included, but no distinctive features are expected to arise. 

The operators in \eq{4f} are similar to those discussed in top-quark condensation  models. In these and other models of composite quarks only operators with vector current structure that is iso- and color singlet are usually considered.
Here the full set of operators are taken into account since their couplings mix in the RG evolution equations. 
Notice however that the RG regime we are interested in is very different because we do not seek to model chiral symmetry breaking and therefore the IR initial conditions for the parameter $\lambda_i$ are taken at  finite values and no phase transition is present. A discussion of the four-fermion lagrangian in \eq{4f}, and its FP, as a model of chiral symmetry breaking can be found in \cite{Gies2}. The special role played in that context by the operator associated to $\lambda_1$ has been recently emphasized in~\cite{LV}.

We write  the system of coupled $\beta$-function equations for the dimensionless variables $\tilde f$, $h$ and $\tilde \lambda_i = \lambda_i k^2$. In these variables, we find the following RG equations:
\bea
\label{betaf}
\frac{d\tilde f}{dt}&=&\tilde f
-\frac{1}{32\pi^2}\tilde f^3
+\frac{N_c}{4\pi^2}h^2\tilde f \nn \\
\label{betah}
\frac{d h}{d t} & = & 
\frac{1}{16\pi^2}\left[4N_c-3 +\frac{16}{\tilde f^2}
(N_c \tilde\lambda_1+\tilde \lambda_2)\right]h^3
+\frac{1}{64\pi^2} \left[\tilde f^2-16(N_c \tilde\lambda_1+\tilde \lambda_2) \right] h \nn \\
\label{betal1}
\frac{d \tilde\lambda_1}{d t} & = & 2\tilde\lambda_1 - \frac{1}{4 \pi^2} \left[
  N_c \tilde\lambda_1^2 +\frac{3}{2} \tilde\lambda_1\tilde\lambda_2-2 \tilde\lambda_1\tilde\lambda_3 -4 \tilde\lambda_1\tilde\lambda_4
\right] \nn \\
\label{betal2}
\frac{d \tilde\lambda_2}{d t} & = & 2\tilde\lambda_2 + \frac{1}{4 \pi^2} \left[
\frac{1}{4} \tilde\lambda_1^2 + 4\tilde\lambda_1\tilde\lambda_3
+ 2 \tilde\lambda_1\tilde\lambda_4
- \frac{3}{4} \tilde\lambda_2^2 + 2(2 N_c+1) \tilde\lambda_2\tilde\lambda_3 + 2 (N_c+2) \tilde\lambda_2\tilde\lambda_4 \right]\\
\label{betal3}
\frac{d \tilde\lambda_3}{d t} & = & 2\tilde\lambda_3 + \frac{1}{4 \pi^2} \left[
 \frac{1}{4} \tilde\lambda_1\tilde\lambda_2 + \frac{N_c}{8} \tilde\lambda_2^2 
+(2 N_c -1) \tilde\lambda_3^2  
+2 (N_c+2) \tilde\lambda_3 \tilde\lambda_4 
- 2 \tilde\lambda_4^2\right] \nn \\
\label{betal4}
\frac{d \tilde\lambda_4}{d t} & = & 2\tilde\lambda_4 + \frac{1}{4 \pi^2} \left[
\frac{1}{8} \tilde\lambda_1^2 -4 \tilde\lambda_3\tilde\lambda_4+ (N_c+2) \tilde\lambda_4^2  
\right]\,. \nn
\eea 

In the RG equations above for the coefficients $\tilde \lambda_i$ we have neglected contributions coming from the Yukawa terms which are proportional to $h^2\tilde f^2$, $h^2 \tilde \lambda_i$ or $h^2 \tilde \lambda_i^2 /\tilde f^2$.
These latter terms are negligible in the UV because we will select fixed points for which the Yukawa coupling approaches zero. They are also subdominant in the IR because the couplings $\lambda_i^2,\tilde f^2$ are IR-free, and such that $\lambda_i^2 /\tilde f^2\rightarrow0$ at low scales. 

\begin{center}
\begin{table}[h!]
\begin{tabular}{|c||c|c|c|c|c|}
\hline
 &  $\tilde \lambda_1$ & $\tilde\lambda_2$& $\tilde\lambda_3$& $\tilde\lambda_4$&$\epsilon_h$\\
\hline
\hline
\textsf{fp0} &  $0$   & $0$ & $0$ & $0$ & $0.5$\\
\hline
\textsf{fp1a} &  $0$   & $-28.71$ & $-7.18$ & $0$ & $1.22$\\
\hline
\textsf{fp1b} &  $0$   & $0$ & $7.85$ & $-9.51$ & $0.5$ \\
\hline
\hline
\textsf{fp1c} &   $0$ & $25.61$ & $-4.27$ & $0$ & $-0.15$ \\
\hline
\hline
\textsf{fp1d} &  $25.80$ & $-1.77$ & $0.19$ & $-1.15$ & $-1.42$ \\
\hline
\textsf{fp2a} &  $13.41$ & $20.10$ & $-3.80$ & $-0.24$ & $-1.03$ \\
\hline
\textsf{fp2b} &  $20.86$ & $-3.56$ & $7.04$ & $-8.94$ & $-1.00$ \\
\hline
\textsf{fp2c} &   $0$  & $-36.55$ & $2.34$ & $-13.92$ & $1.43$ \\
\hline
\textsf{fp2d} &   $0$  & $0$ & $-15.79$ & $0$ & $0.5$ \\
\hline
\textsf{fp2e} &    $37.17$ & $-37.36$ & $-8.43$ & $-1.65$ & $-1.38$ \\
\hline
\textsf{fp2f} &  $-2.92$ & $32.59$ & $4.67$ & $-12.04$ & $-0.10$ \\
\hline
\textsf{fp3a} &    $0.$ & $31.67$ & $4.67$ & $-12.06$ & $-0.30$ \\
\hline
\textsf{fp3b} &   $19.95$ & $-8.59$ & $-15.27$ & $-0.36$ & $-0.80$ \\
\hline
\textsf{fp3c} &   $31.22$ & $-44.52$ & $0.73$ & $-13.38$ & $-0.74$ \\
\hline
\textsf{fp3d} &  $-4.87$   & $1.54$ & $-5.42$ & $-20.10$ & $0.83$ \\
\hline
\textsf{fp4} &   $0$  & $0$ & $-5.42$ & $-20.13$ & $0.5$ \\
\hline
\end{tabular}
\caption{Values of the coefficients  $\tilde \lambda_{i*}$ for the 16 FPs discussed in the text. $\tilde f_* = 17.78$, $h_*=0$ for all FPs. The FP \textsf{fp1c} is boxed.}
\label{fps}
\end{table}
\end{center}

Notice that only the operators proportional to $\lambda_1$ and $\lambda_2$ contribute to the $\beta$-function for $h$. The other two operators do not contribute because of their chiral properties. Moreover,  it is crucial that the combination 
$16(N_c \tilde\lambda_1+\tilde \lambda_2)$  at the FP of $\tilde \lambda_1$ and $\tilde \lambda_2$  be different from zero and larger than $\tilde f^2$ because otherwise there would be no UV FP.

\begin{figure}[h]
\begin{center}
\includegraphics[width=4in]{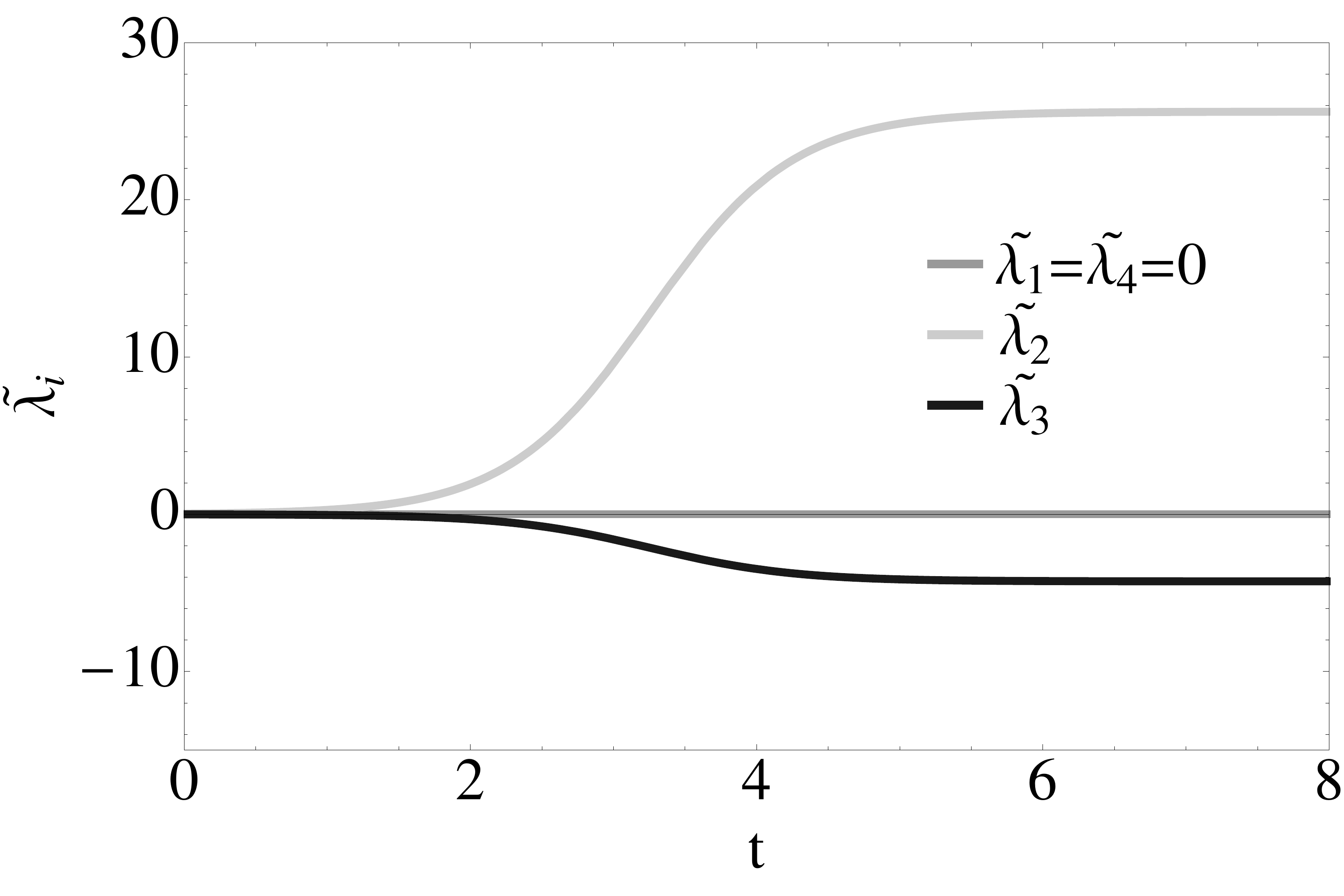} 
\caption{\small Running of the $\tilde\lambda_i$ for the FP \textsf{fp1c}. $\tilde \lambda_1$ and $\tilde \lambda_4$ are equal to zero at all energies. \label{lambdas}}
\end{center}
\end{figure}

From here on we fix $N_c=3$.
We begin by observing that the $\beta$-functions of the $\tilde\lambda_i$ form a closed sub-system. The numerical study of these equations reveals the presence of 16 real
fixed points. Their coordinates are given in the first four columns in Table I.
They are listed in order of decreasing trace of the stability matrix 
$\frac{\partial\beta_{\tilde\lambda_i}}{\partial\tilde\lambda_j}$,
from the most UV-repulsive, the Gaussian FP \textsf{fp0}, to the most UV attractive \textsf{fp4}.  The coefficients of the operators related to UV-repulsive directions are completely determined by the AS condition. Hence, the FPs with a low number of UV-attractive directions are the most predictive ones, and hence the most phenomenologically appealing. The number in the name of the FP is the number
of relevant (UV attractive) directions.
These values can then be used to find the zeroes of $\beta_h$, and these
in turn are used to find the zeroes of $\beta_{\tilde f}$.

We are interested in the sixteen FPs for the complete system with $h_*=0$; this requirement implies $\tilde f_*^2=32\pi^2$.
At each of these sixteen FPs  the direction $\tilde f$ is always a relevant one, with eigenvalue $-0.45$; the direction $h$ is also an eigendirection, with eigenvalue
\be
\epsilon_h=\frac{\partial\beta_h}{\partial h}\Bigg|_*
=\frac{1}{64\pi^2}\left(\tilde f_*^2
-16(N_c \tilde\lambda_{1*}+\tilde \lambda_{2*})\right)\,.
\ee

The numerical values of $\epsilon_h$ are listed in the last column of Table I.
The seven FPs with $\epsilon_h>0$ are physically uninteresting since the requirement of flowing to one of them in the UV
implies that $h(t)=0$ at all scales.
The other nine FPs of the fermionic sector for which $\epsilon_h<0$ also admit a FP with nonzero $h$.
We are not interested here in these additional nontrivial FPs,
but restrict our attention to those with $h_*=0$.
For the $\tilde\lambda_i$, the condition of flowing to \textsf{fpnx} in the UV yields
$4-n$ predictions. The values of $\tilde f$ and $h$ remain always
free parameters, to be fixed by comparison with the experiment.

We have studied numerically the trajectories emerging from the FPs
in the directions of the relevant eigenvectors.
Some of them lead to divergences, others flow to other FPs.
Here we shall consider only the single renormalizable trajectory
that ends at \textsf{fp1c} in the UV. This is a natural choice as it is the most predictive (the one with the smallest number of relevant directions) among the FPs with $\epsilon_h<0$.

In order to select the fine-tuned initial conditions in the IR
that guarantee AS at \textsf{fp1c}, we have first solved numerically the flow equations
of the fermionic subsystem for decreasing $t$, starting at an initial point
$\tilde\lambda_{i*}+10^{-8}v_i$, where $v_i$ is the relevant eigenvector.
This trajectory is attracted towards \textsf{fp0} after roughly 20 $e$-foldings,
so the four-fermion couplings can be taken arbitrarily small by selecting 
the IR value of $t$ appropriately.
We then shift $t$ such that this value is zero, in accordance with our
convention that $t=0$ corresponds to the scale $k_0=\upsilon$.
Now we pick a trajectory for the whole system by fixing the initial
values of the $\tilde\lambda_i$ to agree with the ones find by this method,
while the initial value of $\tilde f$ is $2k_0/\upsilon=2$
and the initial value for $h$ is $0.7$.
These agree with the initial values of the trajectory discussed in section II.

The result is shown in Figs.~\ref{fermions} (continuous curves) and \ref{lambdas}.
We see that for small $t$ the couplings $\tilde f$ and $h$ behave
as in the model of section II, with $\tilde f$ and $h$ both increasing.
At some point, however, $\tilde\lambda_2$ becomes sizable and then the
last term in the right hand side of the second equation in (\ref{betah}) pulls $h$ towards zero.
The trajectory is therefore characterized by a crossover from the
IR regime where the fermionic interactions are mainly of Yukawa type,
with IR free four-fermion interactions,
and the UV regime dominated by the contact interactions,
with UV free Yukawa coupling.
This is similar to the behavior discussed in \cite{schwindt},
except that here we do not bosonize the contact interactions:
we see that in the presence of the Goldstone bosons the
crossover is automatic.

\section{Experimental constraints}

In order to select a realistic trajectory among the various possibilities
one would need to compare with experimental results. The model is  consistent with EW precision measurements, as discussed in~\cite{noi-again}, where the values of the EW parameters $S$ and $T$ are computed for the gauged NL$\sigma$M; the addition of the fermion contact interactions does not modify this result.

\begin{figure}[h]
\begin{center}
\includegraphics[width=4in]{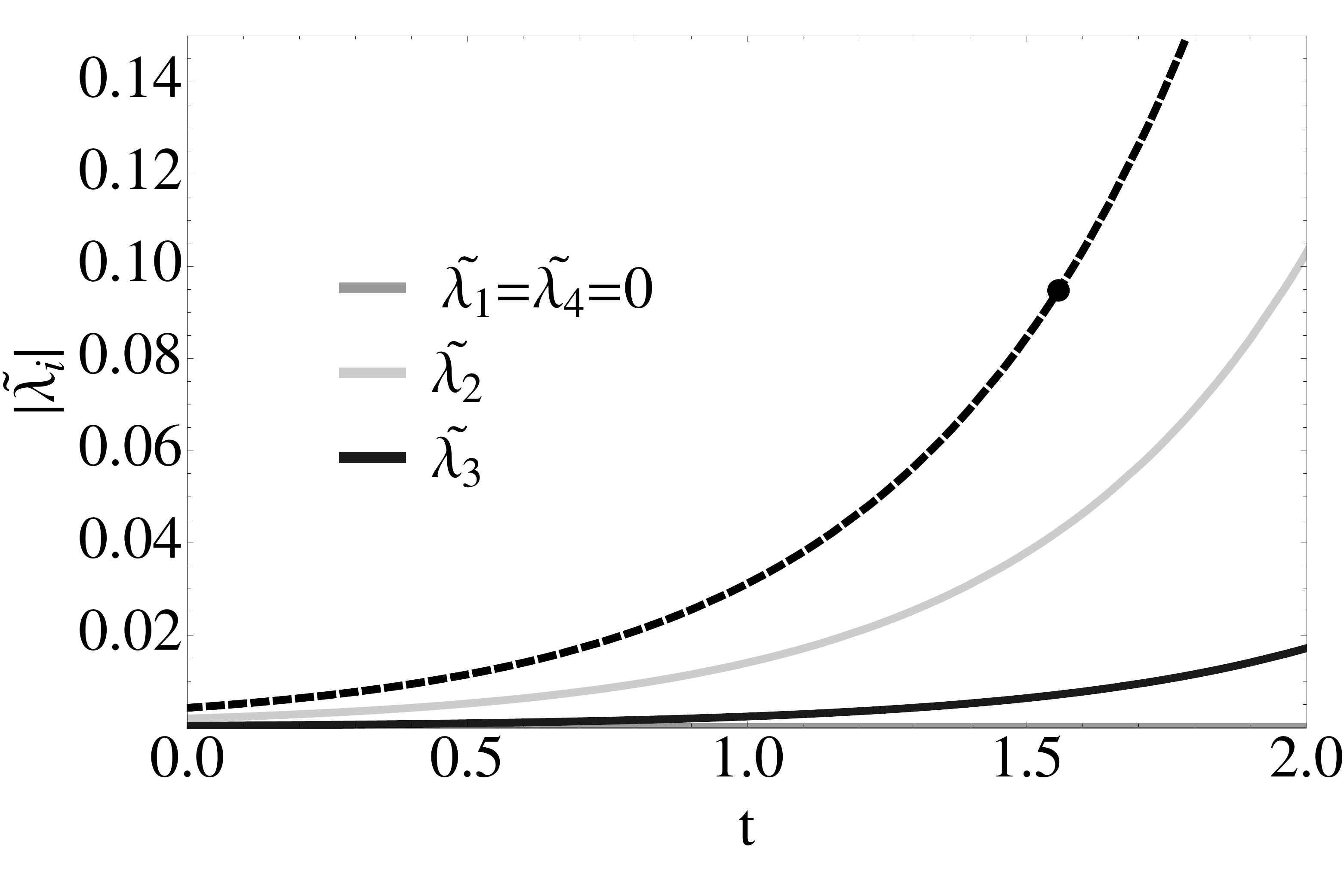} 
\caption{\small A zoom at low-energy of Fig.~\ref{lambdas} for the values (continuous curves) of the absolute values of the four-fermion operator coefficients $\tilde \lambda_i$. The dashed curve is the corresponding size for the contact interaction for the bound at $\Lambda = 9.5$ TeV. The dot at $t=1.55$ on the dashed curve represents the bound at the reference value of the quark-level effective energy of 1.17 TeV. The values of the $\tilde \lambda_i$ must lie below the dashed curve for the experimental bound to be satisfied. \label{exp}}
\end{center}
\end{figure}

Unfortunately the current bounds on contact interactions have been published only for the case in which a single operator, namely that proportional to our $\lambda_3$ is present~\cite{Eichten}. This is rather unrealistic, given the RG mixing, but no need for a more detailed study was felt necessary. 

In this section we aim to show how the experimental bounds are in principle already able to tell us something about the size of the new interactions. Here we  take, conservatively but rather unrealistically, the current experimental bound on $\lambda_3$ and enforce it on all coefficients as if they were contributing in the same manner to the partonic cross sections. In addition we assume that all three generations contribute with identical coefficients $\tilde\lambda_i$. A full Montecarlo simulation, and more comprehensive bounds, will be presented in another publication~\cite{noi}. 

The experimental bound is a lower bound of the  so-called contact interaction scale $\Lambda$ which is related to the coefficient $\tilde \lambda_3$ by the identity
\be
\tilde \lambda_3(k) = \frac{2 \pi}{\Lambda^2} k^2 \, . \label{10}
\ee
A bound on the value of $\Lambda$ translates into a curve in the $k,\tilde\lambda_3$ plane because of the energy dependence in \eq{10}. As a crude estimate, we decide to impose the experimental bounds at the quark-level effective energy scale of the LHC, namely $k_{eff} =\sqrt{s}/(2\times 3) = 1.17$ TeV, where the factor of 2 comes from the sharing of the energy with the gluons and the factor of 3 from the assumed equal energy partition among the three valence quarks. As mentioned before, for simplicity we will impose the same constraint to the absolute value of all four coefficients $\tilde \lambda_i$.

As an example of the kind of constrains we can thus obtain, we take the most recent published bound~\cite{best} of $\Lambda = 9.5$ TeV for an integrated luminosity of 36 pb$^{-1}$ and values of $\tilde \lambda_i$ on a RG trajectory leading to
 the  FP discussed in the previous section,  namely \textsf{fp1c}.  Of course, the FPs with fewer relevant deformations lead to more
predictions and therefore will be easier to disprove.
As we can see from Fig.~\ref{exp},  the bound is satisfied.
 A future bound of $\Lambda = 30$ TeV is expected as the integrated luminosity of the LHC come close to 100 fb$^{-1}$~\cite{acosta}.
 
 In comparing with the experimental data, one may worry whether the power-law running of $h$ and $\upsilon$ may imply  potentially important mass corrections to the cross sections. 
 We compared  the RG running of the masses for the solution above and verified that it does not deviate from the logarithmic one of dimensional regularization for more than 10\%, at least below the scale so far explored, namely 600 GeV.



\begin{thebibliography}{99}

\bibitem{Weinberg}
 S.~Weinberg, ``Critical Phenomena For Field Theorists,'' 
 Lectures presented at Int. School of Subnuclear Physics, Ettore Majorana, Erice, Sicily, Jul 23 - Aug 8, 1976.

\bibitem{reviews}
S. Weinberg, In {\it General Relativity: An Einstein centenary survey}, 
ed. S.~W. Hawking and W. Israel, pp.790--831, Cambridge University Press (1979);\\
M. Niedermaier and M. Reuter, Living Rev. Relativity 9, 5  (2006);\\
R. Percacci, 
in ``Approaches to Quantum Gravity: Towards a New Understanding of Space, Time and Matter'' 
ed. D. Oriti, Cambridge University Press (2009). 

\bibitem{zinn}  J. Zinn-Justin, {\it Quantum Field Theory and Critical Phenomena}, Oxford University Press (2002).

\bibitem{codello}
  A.~Codello and R.~Percacci,
  Phys.\ Lett.\   {\bf B672} 280 (2009)
  [arXiv:0810.0715 [hep-th]].
 
\bibitem{fptz}
 M.~Fabbrichesi, R. Percacci, A. Tonero and O. Zanusso,
  Phys.\ Rev.\  {\bf D83}, 025016 (2011).
  [arXiv:1010.0912 [hep-ph]].

\bibitem{hasenfratz}
P. Hasenfratz, Nucl. Phys. {\bf B 321} 139 (1989);\\
R.~Percacci and O.~Zanusso,
Phys. Rev. {\bf D81} 065012 (2010) [arXiv:0910.0851 [hep-th]].
  
\bibitem{ramond} 
 H.~Arason, D.~J.~Castano, B.~Keszthelyi {\it et al.},
  Phys.\ Rev.\  {\bf D46}, 3945-3965 (1992).

\bibitem{Gies}
H. Gies and M.M. Scherer, 
Eur. Phys. J. {\bf C66} 387-402 (2010) 
[arXiv:0901.2459 [hep-th]];\\
H. Gies, S. Rechenberger and M.M. Scherer, 
Eur. Phys. J. {\bf C66} 403-418 (2010) 
[arXiv:0907.0327 [hep-th]].

\bibitem{Gies2}
H.~Gies, J.~Jaeckel, C.~Wetterich,
  Phys.\ Rev.\  {\bf D69}, 105008 (2004)
  [hep-ph/0312034].
  
   \bibitem{LV}
  L.~Vecchi,
  Phys.\ Rev.\  D {\bf 82}, 045013 (2010)
  [arXiv:1004.2063 [hep-th]].

\bibitem{schwindt}
  J.~M.~Schwindt and C.~Wetterich,
  Phys.\ Rev.\  D {\bf 81} (2010) 055005
 [arXiv:0812.4223 [hep-th]].
 
   \bibitem{noi-again} 
   M.~Fabbrichesi, R.~Percacci, A.~Tonero and L.~Vecchi,
  arXiv:1102.2113 [hep-ph].


\bibitem{Eichten}
  E.~Eichten, K.~D.~Lane, M.~E.~Peskin,
  Phys.\ Rev.\ Lett.\  {\bf 50}, 811-814 (1983).

\bibitem{noi} F. Bazzocchi, U. De Sanctis, M. Fabbrichesi and A. Tonero, in preparation.

\bibitem{best} 
G.~Aad {\it et al.} [ATLAS Collaboration ],
   arXiv:1103.3864 [hep-ex].



\bibitem{acosta}
See, for instance: M.~L.~Vazquez Acosta [ATLAS and CMS Collaborations],
   arXiv:0709.2518 [hep-ph].
   
 


\end{thebibliography}
\end{document}